*Editorial: Physical Review C* in its Fortieth Year

This year, 2010, is the 40[th] anniversary of *Physical Review C* as a separate section of the Physical Review. We write here with a double purpose: first, to describe how PRC has evolved and to explore how this evolution has reflected changes in the research style and interests of scientists working in nuclear physics; and second, to solicit suggestions for improvements that will help guide PRC's future evolution.

The founding of PRC was a result of the growth of the physics research enterprise and the expanding size of the published literature. *Physical Review* was one journal from 1893 to 1964, but then began to split into two and eventually 5 parts, and in 1970 became PRA, PRB, PRC, and PRD. PRA was further split into PRA and PRE in 1993. In 1992, PRC published as many pages as all of PR in 1951, and it grew by almost another factor of 2 from 1992 to 2009. This inexorable growth and the increasing costs to libraries for acquisition and shelving were driving forces behind the eventual conversion to online publication and subscriptions. Of total APS subscriptions, 70% are now electronic only, and the fraction is growing at 6–10% per year.

The world total of nuclear research publications, as measured by the number of original articles (not conference proceedings) in the four principal nuclear physics journals, *European Physics Journal A* (*Z. Phys. A*), *Journal of Physics G*, *Nuclear Physics A*, and *Physical Review C*, has not grown, averaging around $1430\pm80$ between 1979 and 2009. However, the fraction of articles published in PRC has grown steadily, from about 33% in 1971, a year after its founding, to 75% in 2009.

One cannot be certain why this growth in PRC occurred, but much of it can be attributed to two factors: the decrease and elimination of author payments that allowed more people to publish in PRC and the growing reputation of the journal. Physical Review had a history of charging authors to help spread the cost of the journal between authors and readers. Levying such "page charges" reduced the burden on libraries and, in principle, provided a wider distribution of information. It is reasonable, moreover, to regard the cost of publication as part of the cost of doing research. In the real world, however, imposition of page charges can have unintended consequences, driven by several factors: that other journals find a competitive advantage in not having page charges; that the cost of publication varies greatly from journal to journal; and that the amount of research support, relative to publication costs, differs for different research areas and for different countries.

The editors of PRC and PRD had abundant first-hand evidence that page charges in an era of declining research support were driving authors to commercial journals that did not have page charges. Some countries, furthermore, were not willing to have their research agencies support page charges. Since PRC and PRD faced strong competition from commercial journals that were much more expensive but did not have page charges, the editors argued that the net result of page charges was to drive papers to these journals, resulting in an overall increase in library costs. Theorists with relatively small grants and relatively large publication rates found it particularly difficult to pay page charges. Arguments by the editors for elimination of page charges for PRC and PRD resulted in an experiment: Beginning in July 1992, page charges in PRC would be waived for papers prepared in an appropriate electronic format. PRC then experienced a jump of over 20% in published papers between 1992 and 1994. At present, a publication charge ($1235 for regular articles) is levied for articles not provided in an acceptable electronic format, but very few manuscripts now incur these charges.

PRC is the smallest of the PRA–PRE journals, a situation with both disadvantages and advantages. Certain economies of scale are absent, and it can be difficult to avoid disruptions associated with personnel departures, but the small size also makes it possible to experiment with changes: PRC was the first section to have a remote (i.e., outside of the APS Editorial Office) editor when Heinz Barschall assumed the editorship in 1972; in July 1989, PRC was the first section to move Rapid Communications to the front of the print journal and increase their length to five pages; PRC led the way in the use of electronically formatted manuscripts, having always had the highest percentage of such submissions; in July 1999, PRC was the first section of PR to go electronic-first with six-digit article identifiers; PRC reduced the length of "Comments" to two pages beginning in July 1996, to encourage better focused Comments and to limit disputes; and PRC publishes links to selected articles from other PR journals that may be of interest to PRC readers. In the near future, PRC will move to the use of "structured abstracts" to provide a more uniform format for abstracts, and increase the efficiency of electronic searching and summary.

The largest changes in the content of the journal reflect changes in research interests and in the ways research is done. To illustrate these facts by example, we compare the number of regular articles for the earliest date when detailed information is available and for 2009. The results are shown in the figures. Large and/or complex facilities are now the rule, resulting in larger research collaborations, larger author lists, and a smaller or at least limited growth in the number of experimental articles. This is evident in the slightly smaller number of US experimental articles in 2009 and the slightly larger number of non-US articles; both changes are not far outside statistical uncertainties. The number of non-US theoretical papers has grown greatly. This and the remarkable growth in the average number of authors are the major changes seen in Fig. 1.

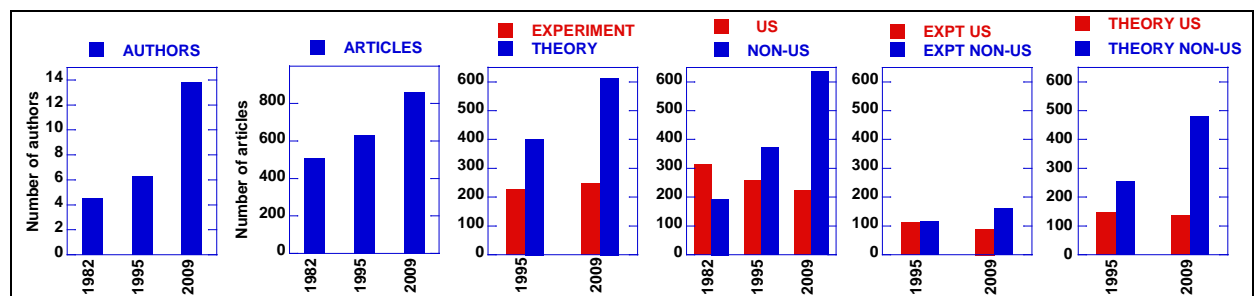

FIG. 1. (Color) *Physical Review C* statistics for various years as shown on the abscissa. The ordinate label of the second pane describes all panes to its right. In all cases the number of *regular* articles is shown.

Figure 2 shows the evolution of article subject matter. The largest percentage growth is in two areas: in Relativistic Collisions and in Nuclear Astrophysics. The classic areas of nuclear physics, Nuclear Structure and Nuclear Reactions, remain strong, but their emphasis has shifted significantly toward nuclei far from stability and studies with radioactive beams.

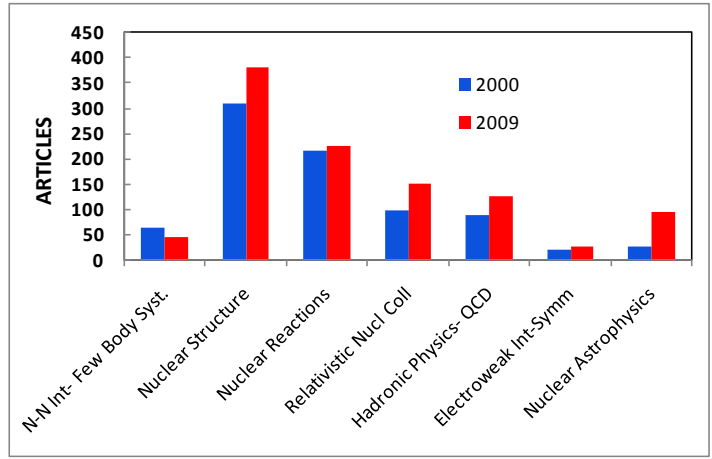

FIG. 2. (Color) Evolution of the size of various sections of *Physical Review C*. All numbers shown are for *all* articles.

A long-standing issue involves research at the border between nuclear and particle physics. As the boundaries of nuclear physics have changed, papers that would once have been classified as high-energy or particle physics now lie in the nuclear physics domain. Yet, the transition is incomplete--authors sometimes submit to their previous journal, and editors are hesitant to recommend a transfer to PRC. We believe that authors would be more uniformly treated and the nuclear science literature would be better accessed if ways were found to complete this transition.

*Physical Review C* and the nuclear physics community owe a debt of gratitude to the many scientists who have served as Associative Editors and Editorial Board Members. Those now on-board are listed at http://forms.aps.org/historic/mast710-prc.pdf. During the early PRC years, there were usually, in addition to staff at Ridge, a remote Editor and a remote Associate Editor, handling experimental and theoretical papers, respectively. As years passed, it became clear that referee selection would be improved and the resulting reports better understood, if wider expertise were available. Additional Associate Editors were recruited; at present there are seven (six remote) handling papers in different areas. A further advantage in the expanded number of handling editors is the additional opportunity this provides for interactions with authors.

We seek your advice and input to help guide the future evolution of PRC. Some possibilities, to stimulate additional input, are the following: Would it be useful to allow authors to cross-list their articles from other PR Journals in PRC, or for PRC authors to list their articles in more than one section of PRC? (It might facilitate simple RSS access of a broader array of physics.) Or would allowing authors to add a single-sentence description of their article to an RSS feed enhance the usefulness of these feeds? Or would a tablet-computer compendium of, say, the last ten years of PRC be useful? These suggestions might not be simple to carry out efficiently, but the editors would appreciate hearing your views and any other suggestions; these can be addressed to the signees, to the Associate Editors, or to prc40th@ridge.aps.org.

Sam M. Austin, Editor 1988–2002; austin@nscl.msu.edu
Benjamin F. Gibson, Editor 2002–present; gibson@aps.org